\PassOptionsToPackage{dvipsnames}{xcolor}

\documentclass[sigconf,screen]{acmart}

\acmYear{2026}
\acmSubmissionID{2316} 
\acmDOI{XXXXXXX.XXXXXXX}
\acmISBN{978-X-XXXX-XXXX-X/26/MM} 
\acmConference[DIS '26]{Designing Interactive Systems}{June 13--17, 2026}{Singapore} 

\setcopyright{none}
\copyrightyear{\textcopyright{} Zeyu Huang, Zhifan Guo, Xingyu Li, Xiaojuan Ma, Noura Howell, 2026. This preprint is for your personal use and not for redistribution. The definitive Version of Record will be published soon in ACM Digital Library in early 2026}

\usepackage{xspace} 
\usepackage{enumitem} 
\usepackage{tabularx} 
\usepackage{multirow} 
\usepackage{subcaption} 
\usepackage{hyperref} 
\usepackage[nameinlink,noabbrev,capitalise]{cleveref} 

\newcommand{\name}{HeartSway\xspace}
\newcommand{\edit}[1]{{#1}\xspace}

\begin{document}
\title[\name]{\name: Exploring Biodata as Poetic Traces in Public Space}


\author{Zeyu Huang}
\authornote{This work was done during the author's research visit to Georgia Institute of Technology.}
\authornote{Both authors contributed equally to this research.}
\email{zhuangbi@connect.ust.hk}
\orcid{0000-0001-8199-071X}
\affiliation{%
  \institution{The Hong Kong University of Science and Technology}%
  \city{New Territories}
  \state{Hong Kong}%
  \country{China}%
}%
\affiliation{%
  \institution{Georgia Institute of Technology}%
  \city{Atlanta}%
  \state{Georgia}%
  \country{USA}%
}

\author{Zhifan Guo}
\authornotemark[2]
\email{zguo381@gatech.edu}
\email{zhifanguo@ucsb.edu}
\orcid{0009-0007-7999-5205}
\affiliation{%
  \institution{Georgia Institute of Technology}%
  \city{Atlanta}%
  \state{Georgia}%
  \country{USA}%
}
\affiliation{%
  \institution{University of California, Santa Barbara}
  \city{Santa Barbara}
  \state{California}
  \country{USA}
}

\author{Xingyu Li}
\email{xingyu@gatech.edu}
\orcid{0000-0002-4456-417X}
\affiliation{%
  \institution{Georgia Institute of Technology}%
  \city{Atlanta}%
  \state{Georgia}%
  \country{USA}%
}

\author{Xiaojuan Ma}
\authornote{Corresponding authors}
\email{mxj@cse.ust.hk}
\orcid{0000-0002-9847-7784}
\affiliation{%
  \institution{The Hong Kong University of Science and Technology}%
  \city{New Territories}
  \state{Hong Kong}%
  \country{China}%
}%

\author{Noura Howell}
\authornotemark[3]
\email{noura.howell@gmail.com}
\orcid{0000-0002-7296-2714}
\affiliation{%
  \institution{Georgia Institute of Technology}%
  \city{Atlanta}%
  \country{USA}%
}
\affiliation{%
  \institution{University of Southern Denmark}%
  \city{Vejle}%
  \country{Denmark}%
}

\renewcommand{\shortauthors}{Huang et al.}

\begin{abstract}
  Human traces scattered across urban landscapes can signify our everyday lives and societal vibrancy in subtle and poetic forms. In this paper, we explore how designed technology can engage biodata as evocative traces. To this end, we present the design, implementation, and evaluation of \name{}, an interactive hammock that captures a user's heart rate and micro-movements as traces and replays them as an embodied experience for the next visitor. Through a qualitative field study (N=10), we find that \name{} evokes feelings of connection, curiosity about prior users, and appreciation for shared human vitality. Our work contributes to understanding anonymous archival biodata as a design material for experiential urban traces. We offer design considerations for intimate asynchronous encounters between strangers in public spaces and for reimagining public amenities.
\end{abstract}



\begin{teaserfigure}
  \centering
  \includegraphics[height=6cm]{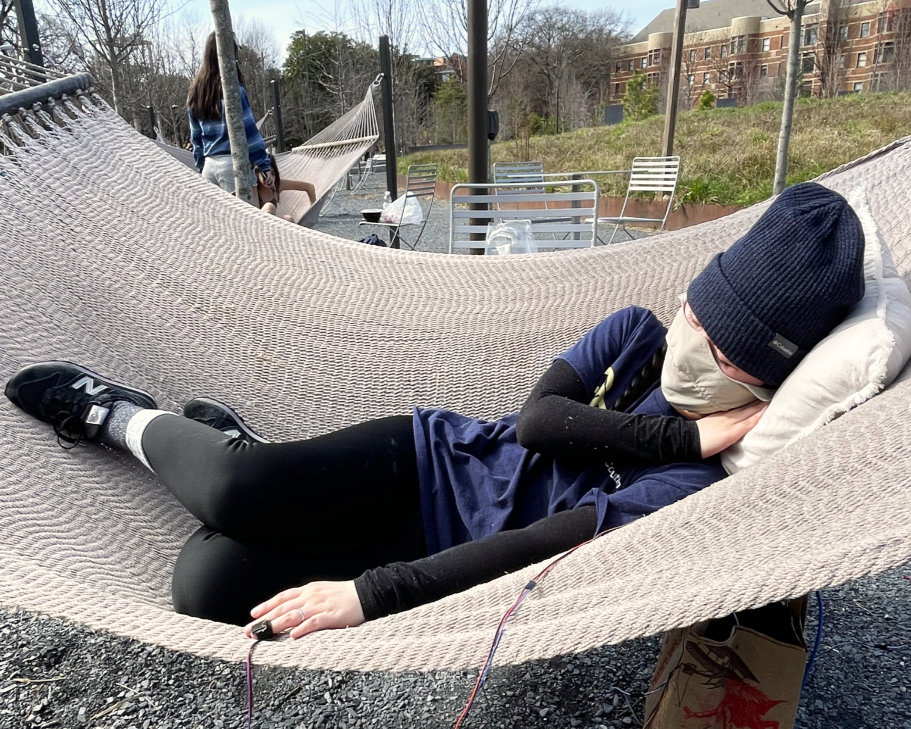}
  \Description{A person lying on a hammock in the public space. Electronic components can be seen around the hammock that are wired to microcontrollers underneath the hammock.}
  \caption{\name captures the biodata of the previous hammock user and embodies it through vibrations and hammock swinging while the next person is lying in it. This way, the archives of biodata series are physicalized yet deprived of contexts, similar to ``traces'' in urban spaces; and the hammock becomes the material that holds these urban dwellers' traces. Participants experienced \name on different days and times in a public garden. Our findings unpack how \name's ambiguous and authentic messages trigger poetic imagination and the revelation of human livingness. }\label{fig:teaser}
\end{teaserfigure}


\maketitle

\section{Introduction}

Traces in urban spaces are quiet markers of human dwelling. From the warmth left on a public bench to the faint impression of footprints, they symbolize the passage of unique lives in our shared neighborhoods~\cite{kaiserDwellingSpeakingUnnoticed1996} and enrich our everyday perception of otherness~\cite{murphyDwellingTogetherObservable2017}.
In contrast to the homogenizing tendencies of modern urban planning, traces are intrinsically personal, injecting dynamic narratives into public settings and celebrating the diversity of human experience~\cite{murphyDwellingTogetherObservable2017}.
With their inherent ambiguity, these material residues can also invite poetic reveries of personal anecdotes and broader societal vitality~\cite{kaiserDwellingSpeakingUnnoticed1996}.

Yet, the city teems with far more traces than those we can physically see or touch.
Our momentary thoughts, affects, and even physiological processes are all organic consequences of our perceptions of and interactions with the surroundings that embed rich latent narratives~\cite{ishiiTangibleBitsSeamless1997}.
However, these signals are doubly elusive: largely imperceptible in the moment and too ephemeral to be preserved in physical form.
In this paper, we are particularly interested in how technology can enable the capture of these latent, invisible traces, as well as engage people in evocative experiences that foreground the poetic and vibrant beauty of traces.

To approach this goal, we turn to biodata---the continuous, involuntary streams of physiological data generated by every human being---as a design material.
\edit{We further deliberately focus on ephemeral biodata signals such as heartbeats and body activities that are typically subconscious and go unnoticed in daily life.}
We investigate \edit{these kinds of} biodata as an alternative form of human trace, one that not only signals presence but also offers a powerful lens into our individuality and affects~\cite{howellEmotionalBiosensingExploring2018}.
Then, we engage in an exploratory design process to craft interactional surrogates that can capture their trace-like essence---subtle, not intruding into regular daily routines, and open to interpretation~\cite{monasteroTracesStudyingPublic2018}.
As a result, we present an installation design named \name, an interactive hammock in public space. It captures a user's heart rate and toss-and-turn movements, and while the next person rests in it, this preserved corporeal imprint is replayed through an embodied experience. The previous user's heart rate is translated into gentle, anchored vibrations, and their restlessness is expressed through the hammock's subtle swinging.
In this way, we render an otherwise intangible presence tangible, inviting each newcomer to inhabit the lingering vitality of a stranger.

We evaluated \name through a qualitative study (N=10) in a real-world outdoor setting on a university campus.
Our findings reveal a range of \edit{interpretations that align with our poetic and affective design goals}, such as spontaneous empathy and curiosity about the previous user, internalization of their physiological state, and appreciation for the world's shared liveliness.
Through this, we open a dialogue on how the body's most intimate, fleeting signals can be reseeded into public space as a form of communal dynamics.
Ultimately, \name invites us to imagine a city that quite literally breathes with the biodata of its inhabitants, transforming urban anonymity into momentary kinship.

In summary, \name offers the following provocations as suggestions for compelling directions in HCI and design:
\begin{itemize}
    \item We advance designing with biodata by showing how its archival, decontextualized display can communicate a sense of poetic vitality among strangers.
    \item We inform designing for interpersonal connection with an approach centered on shared livingness and poetic, anonymous, asynchronous intimacy.
    \item We contribute to designing for public space by transforming single-user public amenities into lenses through which we can see the community's warmth.
\end{itemize}

\section{Related Work}


\subsection{Urban Traces and HCI Design}\label{sec:rw:trace}

In sociology and psychology, researchers celebrate traces in urban spaces for their poetic indication of human activities in the dynamic world. On the one hand, they are generated by individual human beings. The individual perspectives spotlight vivid diversity among the community~\cite{murphyDwellingTogetherObservable2017} and unique socio-emotional narratives that attach observers to the place~\cite{kaiserDwellingSpeakingUnnoticed1996}. On the other hand, their societal and cultural symbolism is largely ambiguous. As the context of the trace's generation is lost in time, observers can derive varied notions from the trace, making the interpretative process poetic~\cite{murphyDwellingTogetherObservable2017}.
The poetic yet sincere characteristics of traces essentially entail a two-fold asynchronous interaction pattern, where traces are not only the result of human activities, but also influencers of every urban dweller upon creation~\cite{kaiserDwellingSpeakingUnnoticed1996}.

In HCI, we are currently witnessing a shift of research focus on traces, from early utilitarian and informative exploitation of traces~\cite{wexelblatFootprintsHistoryrichTools1999} to recent explorations of ubiquitous use in harmony with poetic urban lifestyles~\cite{psyllidisUrbanMediaGeographies2013,hirschUnobtrusiveInterfacesHistorical2020}.
For example, \textit{Jabberwocky} tags both residents' recent trajectories and stationary locations of interests, hinting users of potential re-encounters and social opportunities~\cite{paulosFamiliarStrangerAnxiety2004}; \textit{Optical Stain} prolongs the traces of removed posters on bulletin boards using projectors to advocate serendipitous encountering and discoveries of temporal and social embeddings~\cite{shiraiInteractingInteractionHistories2007}; \textit{The Life of a Small Town} is a collective and ever-growing fabric artwork by a neighborhood to physicalize and commemorate the traces of their residency~\cite{jonesYearInteractionTown2024}; \textit{Traces} projects people's trajectory at the same building on the floor as an unobtrusive and engaging revelation of daily routines, promoting social awareness and serendipity~\cite{monasteroTracesStudyingPublic2018}.

These works show how designers can engage residents by prolonging or spotlighting existing urban traces.
Complementing this, we investigate how designers may alternatively capture new and originally invisible ``traces'', and how the display of otherwise unnoticed and unexpected traces may create new urban interactive experiences with more appreciation for everyday lives.
In this paper, we explore this niche with biodata, for its organic liveliness~\cite{howellLifeAffirmingBiosensingPublic2019,howellEmotionalBiosensingExploring2018,tsaknakiFabulatingBiodataFutures2022,boehnerAffectInformationInteraction2005} and empathetic expressivity~\cite{fengExBreathExploreExpressive2024,aslanPiHeartsResonatingExperiences2020,neidlingerAWElectricThatGave2017} with the individual-oriented and socio-emotional aspects of urban traces.

\subsection{Stranger Interactions}\label{sec:rw:stranger}

Interactions between strangers are a long-standing research topic in HCI, as they contribute to a sense of belonging and community~\cite{simoesaelbrechtFourthPlacesContemporary2016,vanlangeVitaminWhySocial2021, langeOwningCityNew2013}.
For example, some designs like \textit{Jokebox} and \textit{Digital Carpet} engage strangers in playful encounters through synchronized actions~\cite{balestriniJokeboxCoordinatingShared2016,schieckExploringDigitalEncounters2010}.
\textit{AmbiDots} enhance social settings in third spaces with ambient and peripheral interactions~\cite{thompsonAmbiDotsAmbientInterface2022}.
In some other works, public amenities are critically transformed to additionally afford heartfelt interactions of affective communication~\cite{howellLifeAffirmingBiosensingPublic2019} and bodily social play~\cite{patibandaSharedBodilyFusion2024}.

Many designs predominantly focus on synchronous face-to-face encounters, which rely on the user's initiative to engage in prosocial interactions. In reality, urban dwellers often avoid those stimuli that connect each other---the ``asocial'' phenomenon denoted by sociologists as ``civil inattention''~\cite{hirschauerDoingBeingStranger2005} and ``blasé attitude''~\cite{simmelMetropolisMentalLife1950}.
Although some HCI works aim to lower the mental barriers and efforts of shared encounters~\cite{balestriniJokeboxCoordinatingShared2016}, we explore an alternative approach leveraging asynchronous interactions. On such occasions, people do not need direct engagement with unfamiliar others, but they still sense genuine and poetic connections through interacting with the artifacts others have left for them.
\edit{Here, our focus on ``genuine and poetic connection'' is adapted from Stepanova et al.~\cite{stepanovaStrategiesFosteringGenuine2022}: specifically, it is an anonymous, affective, and non-verbal resonance between people rather than an active interpersonal dialogue or the establishment of a collaborative relationship.}

Current asynchronous urban interactions are typically in the form of public displays. From sharing expressive digital content about the neighborhood~\cite{liuPinsightNovelWay2018,rydingLYDSPORURBANSOUND2023} to playful interactions with multimedia artifacts originated from the community~\cite{peltonenItsMineDont2008}, they primarily seek connections between individuals and the community at large.
There remains a significant opportunity to explore more intimate, person-to-person asynchronous interactions---the interaction mode that urban traces usually entail (\cref{sec:rw:trace}).

\subsection{Communicating Biodata Socially}

Biodata is not only capable of communicating health and physiological information. With the HCI community's continuous exploration, an alternative paradigm has grown increasingly popular in recent years: to leverage biodata in social contexts and social relationship support~\cite{mogeSharedUserInterfaces2022}. 
According to Moge et al., most of these works are ``synchronous'', meaning the interactions unfold when both parties are co-situated in the same context and exchange real-time biodata~\cite{mogeSharedUserInterfaces2022}.
For example, hearing the other party's heartbeat sounds during VR communication~\cite{janssenIntimateHeartbeatsOpportunities2010}, and wearing accessories that visualize breaths, heartbeats, and physiological synchrony during face-to-face interactions~\cite{freyBreezeSharingBiofeedback2018, fengExBreathExploreExpressive2024, ozcanMultisensoryWearableBiofeedback2023}.
While these works have shown the social capability of biodata sharing, they do not cover all social scenarios in our daily lives. It is yet to be fully understood how biodata sharing can \edit{influence} our ``asynchronous'' communication---where a series of \emph{past} biodata are told.  


We draw inspiration from several recent pioneer designs about asynchronous biodata communication.
Some works show biodata's capability of stimulating awareness of physiology and augmenting other information channels~\cite{mogeSharedUserInterfaces2022}.
For example, Huang et al.\@ highlighted the audience's frisson moments when watching the same online video~\cite{huangSharingFrissonsOnline2024}, and Hassib et al.\@ annotated speech bubbles with emotion indicators~\cite{hassibHeartChatHeartRate2017}.
Meanwhile, some other works tried to support positive human nature and artistic expression, incorporating biodata series into self-expressive clothes~\cite{jonesWearYourHeart2023}, accessories~\cite{walczakExpreSenseDesigningInteractive2025}, and provocative art pieces~\cite{stamatoMessageBottleInvestigating2024}.
To complement these studies, we aim to inquire into the innate qualities of human biodata and how they can be communicated.
Hence, we focus on another scenario, where biodata is primarily communicated, \edit{perceived}, and understood (cf. self-expression) and as the only information being conveyed (cf. augmenting other information channels).
\edit{Meanwhile, we must also acknowledge that direct biodata sharing carries extra risks beyond intimacy, including surveillance, biometric policing, and identity exposure~\cite{ahmedPrivacySecuritySurveillance2017}. These risks need to be carefully addressed during the exploration.}

\edit{Therefore, we eventually specified our aim to} convey a poetic and authentic feeling of ``otherness''.
As articulated by Hassenzahl et al., it involves cognitive awareness of another person, deliberate ambiguity~\cite{hassenzahlAllYouNeed2012}.
It is widely explored in synchronous biodata sharing designs, including rings~\cite{wernerUnitedpulseFeelingYour2008}, sofas~\cite{sunBreathBePerceived2017}, photo frames~\cite{kimBreathingFrameInflatableFrame2015}, and hugging cushions~\cite{nunezEffectSocialConnectedness2019}.
\edit{And these designs often involve non-numeric, not persistent, unidentified, and non-intrusive representations of biodata, mitigating the risks during direct biodata sharing.}
\edit{Also, we noted that they often leverage haptic feedback design, spanning vibration, pressure, texture, and thermal stimulation, as well as affording a broad range of aesthetic and emotional qualities~\cite{kolerSharingHapticAttributes2020}.}
But fewer designs have attempted to deliver the same feeling through asynchronous communication of past biodata.
Drawing inspiration from the above designs, we aim to explore how \edit{asynchronous biodata expression can foster a similarly poetic feeling of otherness}.



\section{Design of \name}\label{sec:design}

Adopting the reflective design approach, we actively reflected on our own role during the design process, the uncovered design space, and how to support users' self-reflections~\cite{sengersReflectiveDesign2005}.
In particular, we questioned how biodata traces can be embedded in people's everyday living context, and how they can poetically indicate the approachable presence and livingness of another human being.
In this section, to more clearly present our design outcome, we first present the final design artifact in the first subsection. Then, we cover the design process that leads us there in the next subsection. The final subsection covers the technical implementation.

\subsection{Interaction Concept}\label{sec:design:concept}

\name is an interactive hammock with the previous user's biodata presented for the next user to experience.
It is a gentle yet drastic alteration of the functionality of a public hammock, for its preservation of the primary ``lying'' affordance yet breakage of the personal, ``me-time'' experience.
The state of being physically vulnerable and cognitively idle is deliberately leveraged for an unknown stranger's ``friendly approach''.
From the previous user, the hammock has captured the live heart rate and moments of tossing and turning, which are saved in a local database.
When the next user comes, the previous person's heart rate series is reflected in the gentle vibrations of the pillow, and the hammock body swings at the moments when the previous person \edit{moved}.
Meanwhile, the next user's own data is being captured for the yet next user to feel.
These sensory experiences combine spontaneous, uncontrollable data symbolizing physiological activity and controllable motor data indicating bodily vibrancy.
The ambiguous biodata representations convey pure human livingness and proximity of another person, inviting open-ended interpretation instead of predefining any concrete inference from the data~\cite{gaverAmbiguityResourceDesign2003}.
The ambiguity is also strengthened by a unidirectional communication mode, where observers are not able to talk back to the trace creator, leaving a poetic imagination space that mirrors how traces naturally function in the physical world.

\subsection{Design Process}

\begin{figure}[h]
  \centering
  \includegraphics[width=0.95\linewidth]{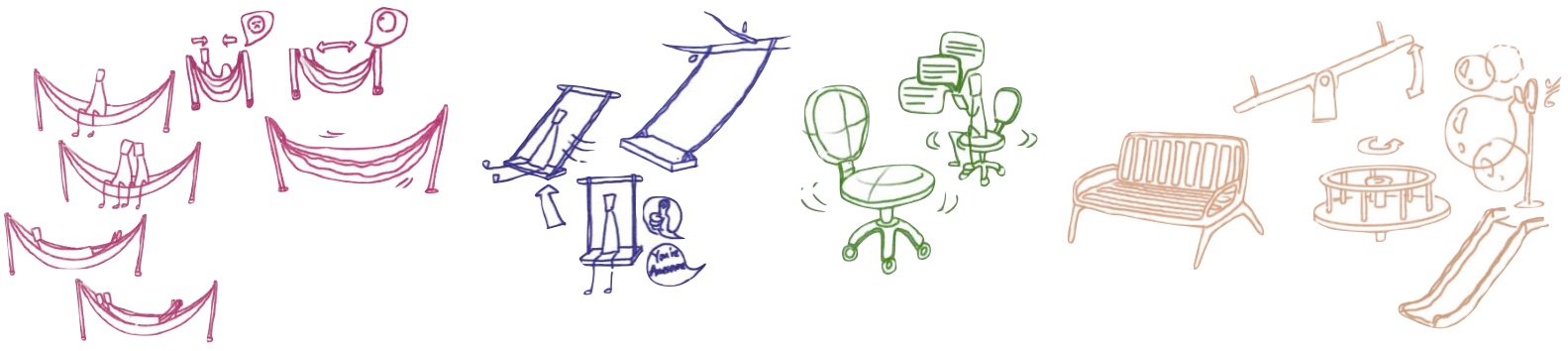}
  \Description{A sketch-style illustration displays various public-space amenities and the movements they afford. The objects include hammocks, swings, chairs, a bench, a seesaw, a merry-go-round, a bubble machine, and a slide.}
  \caption{Our early ideation of tangible artifacts that could embody biodata.}
  \label{fig:design:artifact}
\end{figure}

Drawing from reflective design~\cite{sengersReflectiveDesign2005}, we sought to design an interaction that evokes and prompts reflection on biodata as urban traces with socially connective traits.
First, we anchored our design at public amenities to reinforce the sense of how biodata traces flourish and are embedded in our daily life.
We brainstormed possible amenities to embody intangible biodata and its interactions.
In addition, to make the experience more approachable and empathetic, we primarily searched for those public amenities that offer relaxation or entertainment.
As shown in \cref{fig:design:artifact}, example ideas include hammocks that stretch and contract, swings with varying movement amplitudes, rotating chairs with varying rotation speeds, and bubble wands with varying bubble generation speeds.
In our subsequent discussions, we preferred relaxing amenities over playful ones because they were not limited to stimulating positive emotions; instead, they allowed for more ambiguous exchange and were potentially more suitable for placing users in `listener' roles.
Then, as we decided to deploy the design and conduct studies at Georgia Institute of Technology for practical reasons, we particularly investigated its campus and culture.
\edit{This university has a culturally diverse and respectful community. The campus is also compact with well-established outdoor communal spaces, allowing a vibrant atmosphere and abundant social opportunities. Across the entire campus, hammocks are already installed at several places. They have become a familiar and special part of campus life.}
Thus, combining this situation with the previously mentioned relaxing nature, we selected hammocks as the bearers of biodata recordings.

\begin{figure*}[h]
  \centering
  \begin{subfigure}[t]{0.34\textwidth}
    \centering
    \includegraphics[width=\linewidth]{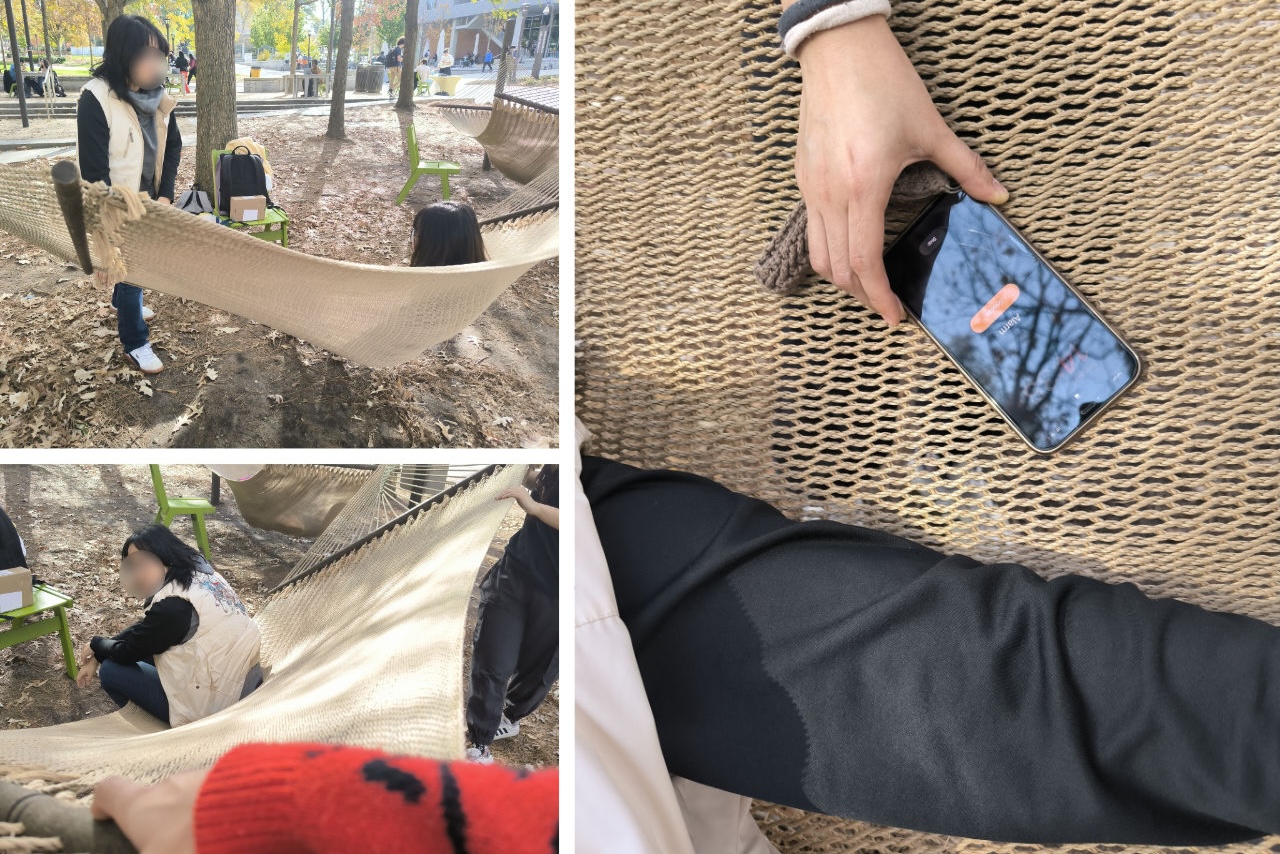}
    \Description{The authors are playing with hammocks and mimicking several haptic experiences on campus.}
    \caption{Through bodystorming, we investigated the feasibility and effects of several haptic modalities when installed in a hammock.}\label{fig:bodystorm}
  \end{subfigure}
  \hspace{\fill}
  \begin{subfigure}[t]{0.3\textwidth}
    \centering
    \includegraphics[width=\linewidth]{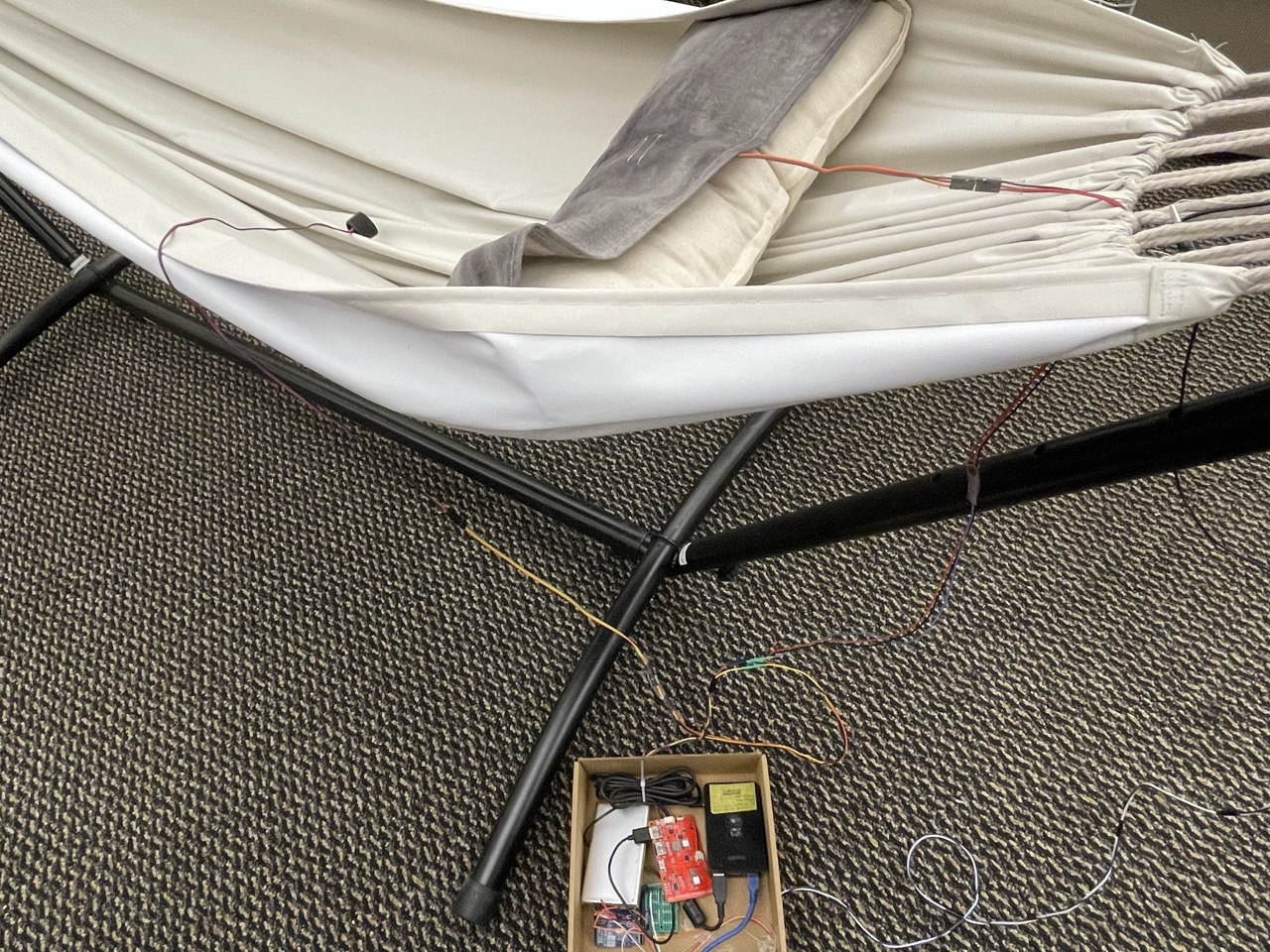}
    \Description{A close-up shot of a hammock with wires attached to it and a box of electronics on the ground.}
    \caption{We iteratively adjusted the wiring and tested to minimize obstruction and ensure functional stability.}
  \end{subfigure}
  \hspace{\fill}
  \begin{subfigure}[t]{0.34\textwidth}
    \centering
    \includegraphics[width=\linewidth]{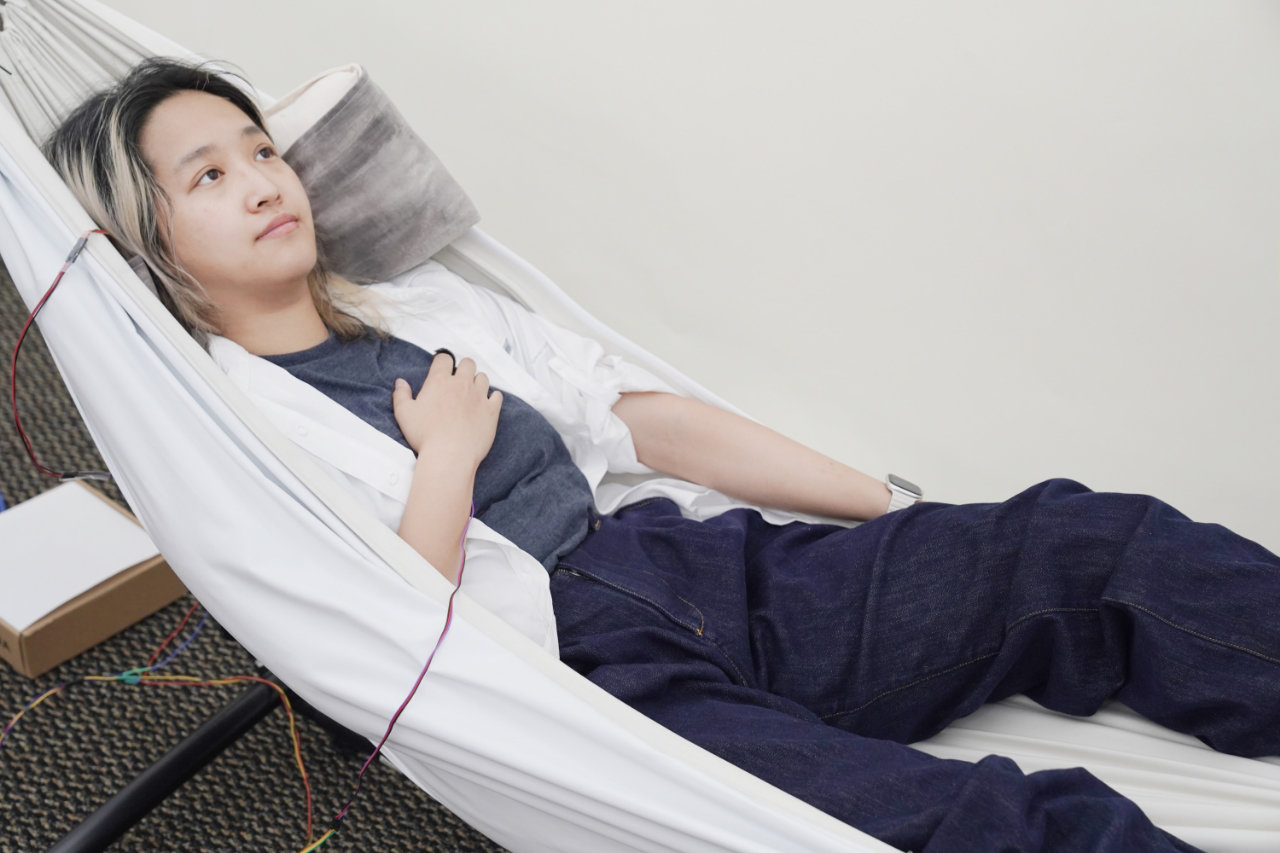}
    \Description{A person lying in the hammock, with her head on a wired pillow and her hand at her chest.}
    \caption{A use scenario of \name. The electronic components are mostly put away from the user, not obstructing their interactions.}
  \end{subfigure}
  \caption{Our explorations of the biodata types and possible interactions in a hammock.}
\end{figure*}

As the second step, we brainstormed the types of biodata and their interactions that might fit, and further tested these ideas via outdoor bodystorming (\cref{fig:bodystorm}) and pilot studies.
We particularly focused on body temperature, heart rate, breathing, and body movements. The motivations behind each choice and our findings are explained below, respectively.
We initially put considerable effort into body heat and thermal materials, following related designs that turn these materials into affective and playful interactions~\cite{choThermoPlayExploringPlayful2024,umairThermoPixelsToolkitPersonalizing2020}. However, during the bodystorming, thermal feedback in hammocks often made us think of complimentary heating blanket without further affective or social indication.
We presumed that this feedback modality might not suit the hammock setting.
While body heat could be alternatively mapped visually with thermochromic and photochromic materials (similar to~\cite{jinPhotoChromeleonReProgrammableMultiColor2019,friskChromaNailsReProgrammableMultiColored2023}), these visual patterns on a hammock were not readily observable to someone lying on their back in the hammock, looking up at the sky, or closing their eyes.
Another major focus during the design process was breaths, because it was also strongly related to affective states and external stimuli, and its perception could elicit intimacy~\cite{haynesJustBreathAway2024}. They could be presented by an inflating and deflating cushion (e.g.,~\cite{haynesJustBreathAway2024}) or the sound of such physiological actions (e.g.,~\cite{howellLifeAffirmingBiosensingPublic2019}). However, breaths turned out to be difficult to sense in the outdoor hammock scenario: body straps were inconvenient to wear, and audio-based methods required complex denoising.
Eventually, we selected heart rates and toss-and-turn body movements as the data of interest.
\edit{In making this selection, we reflected on both their implications and technical feasibility in a hammock.}

The heart rate is a common signal among biodata-related interactions~\cite{wernerUnitedpulseFeelingYour2008,liuAnimoSharingBiosignals2019,aslanPiHeartsResonatingExperiences2020,hassibHeartChatHeartRate2017}.
\edit{It is largely beyond users' conscious control, and the data capture is continuous from the moment someone lies down.
This nature makes it an ambiguous yet evocative signal, as people may bring intuitive understandings of its varied meanings (e.g., heart beating fast due to intense emotion).
We further noticed that vibration is an intuitive feedback modality for heart rates~\cite{yangCraftingRemoteIntimacy2026,wernerUnitedpulseFeelingYour2008}; and it generally supports vast aesthetic experiences like presence~\cite{deyEffectsSharingRealTime2018,liuAnimoSharingBiosignals2019}, intimacy~\cite{hayesAestheticsTouch2017,yangCraftingRemoteIntimacy2026}, and physiological attunement~\cite{morrisonVibrotactileVibroacousticInterventions2018}.
We wanted to create a similar experience of physiological presence and gentle attention to an unknown stranger.
Thus, we experimented with several locations and carriers for the heart rate's vibration feedback, including gloves with embedded electronic components, soft handheld objects, and cuddle cushions. However, these options either imposed a restricted hand position on participants or are difficult to dismantle and clean after continuous shared usage.
We ultimately designed vibration in an outdoor pillow, a common hammock accessory, near the head and neck, where participants could both feel and faintly hear the heartbeat.
}

Toss-and-turn movements are specifically relevant to the hammock setting and might help depict the dynamism of a person in a hammock.
We recorded the exact moments when the user tossed and turned the body, and let the hammock swing itself at the same moments during the next user's stay.
\edit{These movements are conscious actions, providing another perspective of the previous user than involuntary bodily vitality.
They can be easily captured by the changes in the stretchedness of hammock strings.
We explored various mechanical designs of hammock strings and poles to potentially recreate the exact hammock movements at those moments. However, most of them required sophisticated hardware setups, physical modeling, and limited adaptability across different hammocks.
We eventually chose to simply swing the hammock sideways, as it struck a balance between implementation difficulty and signaling human movements.
}

\subsection{Technical Implementation}\label{sec:design:implementation}

\begin{figure}[h]
  \centering
  \includegraphics[width=0.85\linewidth]{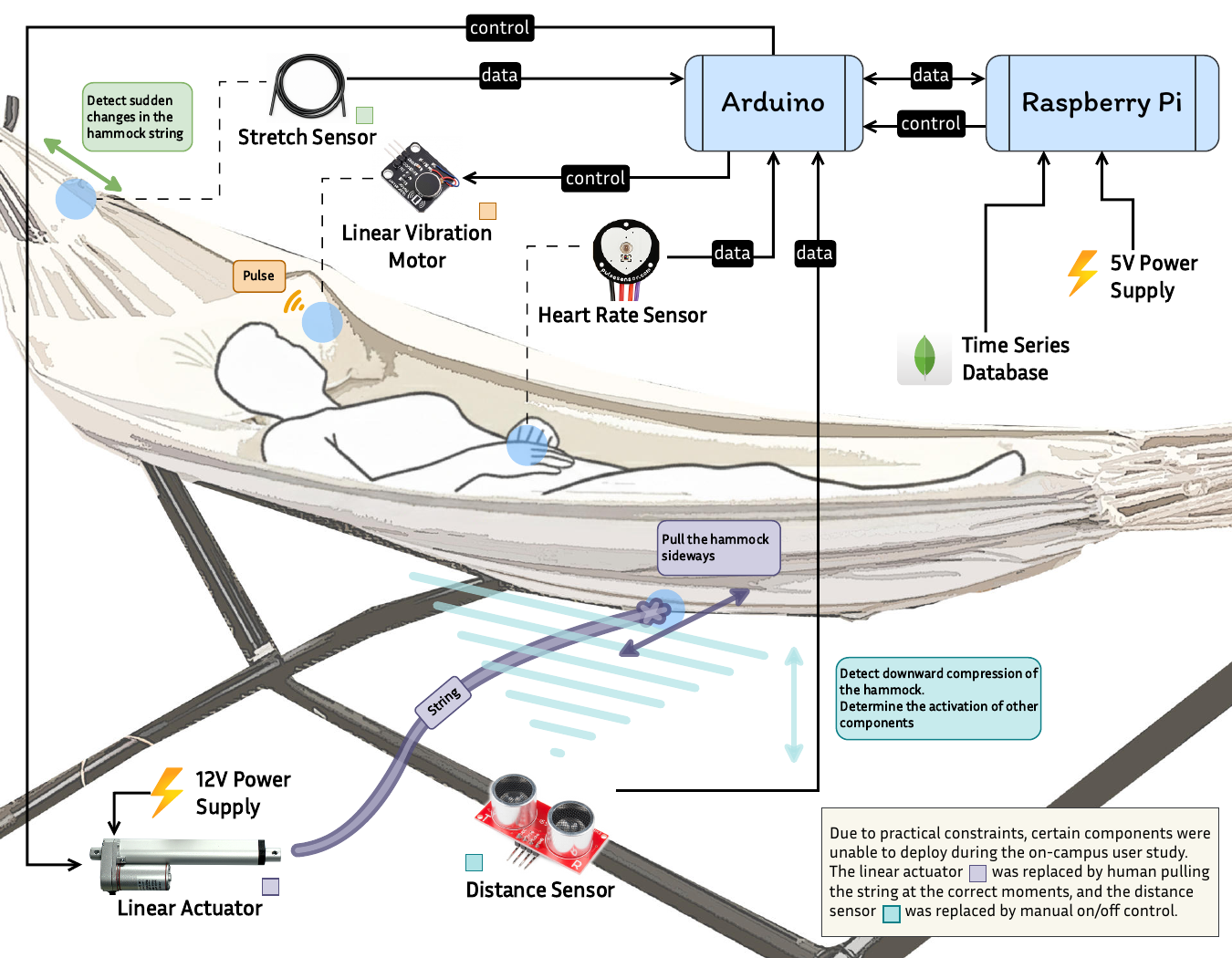}
  \Description{A diagram illustrating the \name system. A hammock is connected to several electronic components: a stretch sensor, a linear vibration motor, a heart rate sensor, a linear actuator through a string, an Arduino receiving data from the sensors and sending control signals to the vibration motor and the linear actuator, and a Raspberry Pi that receives sensor data from the Arduino, forwards the control signals to the Arduino, and stores data in a time series database. Two power supplies are connected to the Raspberry Pi and the linear actuator, respectively.}
  \caption{A diagram of the \name system.}
  \label{fig:system}
\end{figure}

The implementation is shown in \cref{fig:system}, \edit{and it is open-sourced at \url{https://github.com/fhfuih/HeartSway/}.}
A Raspberry Pi and an Arduino coordinate all the electronic components and manage the data, connected via USB for both power supply and data transmission.
The body-moving moments are detected by an Adafruit Conductive Rubber Cord Stretch Sensor\footnote{\url{https://www.adafruit.com/product/519}} attached to the hammock string. Together with the heart rate from a fingertip PulseSensor\footnote{\url{https://pulsesensor.com/}}, these data are sent to the Raspberry Pi's local time series database.
\edit{For both types of data, we only recorded the timing of occurrences, but not their intensities.}
Meanwhile, the previous user's data is loaded from the database, controlling a 5V linear vibration motor in the hammock pillow and a 12V linear actuator that pulls the bottom center of the hammock fabric sideways through a long string.
\edit{The vibration motor was PWM-modulated with a rated speed of 9,000 rpm. Whenever a heartbeat feedback is triggered, the motor vibrated at 40\% strength for 0.1s. The authors tested this configuration and confirmed that it resembled heartbeats with moderate intensity.}
An additional HC-SR04 distance sensor was used to detect human presence. It was placed underneath the hammock facing upward, capturing the periods when the hammock fabric is pushed down by a user.
The Raspberry Pi and Arduino only require a 5V \edit{3A} power supply. Hence, they are powered by a portable \edit{Uninterruptible Power Supply (UPS)} board and a lithium battery. The linear actuator required a dedicated 12V power supply.
Only the human presence (distance) sensing routine is continuously executed. All other sensors and actuators are not activated in the event loop until human presence is detected.

Concerning the data processing, we calculated the Inter-Beat-Interval (IBI) of every beat through the Beats-Per-Minute (BPM) series from PulseSensor.
The body movements are acquired from the stretch sensor's resistance, sampled at 1 Hz.
To extract body-moving moments from the stretchedness series, we needed to identify changepoints from the smoothed series. We first removed outlier data points beyond mean \(\pm\) 3std, where the mean and std were the rolling ones with a window size of 100. Then, we applied a \edit{Pruned Exact Linear Time (PELT)} changepoint detection algorithm with a Gaussian kernel and a penalty of 10, and we recorded the time of all changepoints.
The data for the next user is prepared immediately after the previous user leaves. When the next user comes, the prepared data is sent to the Arduino in a pagination of 30 data points to prevent memory overflow, and it repeats from the start when reaching the end.

We prioritized situating the experience in an open, public place with peaceful surroundings, because the environment that naturally surrounded the hammock was also an integral part of the experience. However, to avoid disrupting the existing campus environment and to comply with site regulations, we could not install large power sources or permanent technical infrastructure.
The site, a quiet park, also did not have existing power sources to support the 12V actuator.
Given these constraints, the distance sensor and the linear actuator were replaced by simpler Wizard-of-Oz~\cite{kelleyEmpiricalMethodologyWriting1983} counterparts.
The Wizard-of-Oz setup allowed us to minimize modifications of the site while still exploring the design in a real-world setting.
Specifically, since the \name hardware was installed and removed at each user study session \edit{with the authors accompanied, the installation and tuning of the distance sensor for human detection became less necessary. Instead, we manually turned the system on and off at every ad-hoc installation with a laptop through SSH under the same local Wi-Fi hotspot.}
For the linear actuator and the swinging feature, we instead plotted the previous user's body-moving moments upon the next person's arrival. One of the authors, being a drummer with an accurate sense of timing, manually pulled the string at the correct moments.
Since the experience is about the previous user's physiological presence, the swinging moments did not need to be perfectly accurate at the millisecond level. Thus, manual simulation was sufficient.
In this deployment, the string was also deliberately made longer, so that this researcher \edit{could remain} 1.5 m away from the hammock so as not to disturb the user.

\section{Study}

\subsection{Participants and Protocol}

We deployed \name on an existing public hammock at a local university.
With \edit{Institutional Review Board (IRB)} approval, we recruited 10 students at the same university via social media\edit{, flyers, } and word of mouth.
\edit{The final participant group contained the design teams' friends, acquaintances, and \edit{extended social networks}. During the recruitment, the authors strove to seek people from different social circles across the university to avoid homogenized social preferences or personalities.}
They are coded A1--A10, and their demographics are listed in \cref{tab:demographics}.

\begin{table}[ht]
  \caption{User study participants' demographics.}\label{tab:demographics}
  \begin{tabular}{llp{1.8cm}p{3.5cm}}
    \toprule
    ID & Age & Gender & Background \\
    \midrule
    A1  & 27 & Female & Education \& AI Literacy \\
    A2  & 33 & Genderqueer (they/she) & User Experience \& Design Theories \\
    A3  & 26 & Female & Digital Media \\
    A4  & 25 & Female & Mechanical Engineering \& Design \\ 
    A5  & 23 & Female & Information \\ 
    \bottomrule
  \end{tabular}
  \hspace{1cm}
  \begin{tabular}{llp{1.8cm}p{3.5cm}}
    \toprule
    ID & Age & Gender & Background \\
    \midrule
    A6  & 26 & Female & Industrial Design \\ 
    A7  & 18 & Female & Industrial Design \\ 
    A8  & 28 & Male & Digital Media \& Education  \\ 
    A9  & 21 & Non-binary / third gender & Computer Science \\ 
    A10 & 30 & Male & Industrial Design \\ 
    \bottomrule
  \end{tabular}
\end{table}

Before the study sessions commenced, we had recorded the lead author's own biodata series as the upcoming experience of the first participant. Then, every participant's biodata was recorded and used as the next participant's experience, as explained in the previous section.

During the user study, a participant was invited to the installation individually and briefed on the design concept, the features, and the meanings of the haptic stimuli.
Then, the participant lay down in the hammock with the fingertip sensor attached for 10 minutes. During the 10 minutes, we suggested that the participant behave freely, following their own habits and understandings. The participant was also allowed to freely adjust the pillow, play with it, or place it elsewhere.
Finally, the participant completed a semi-structured interview, which is described below.

\subsection{Data Collection and Analysis}

\begin{figure*}[t]
  \centering
  \includegraphics[width=\textwidth]{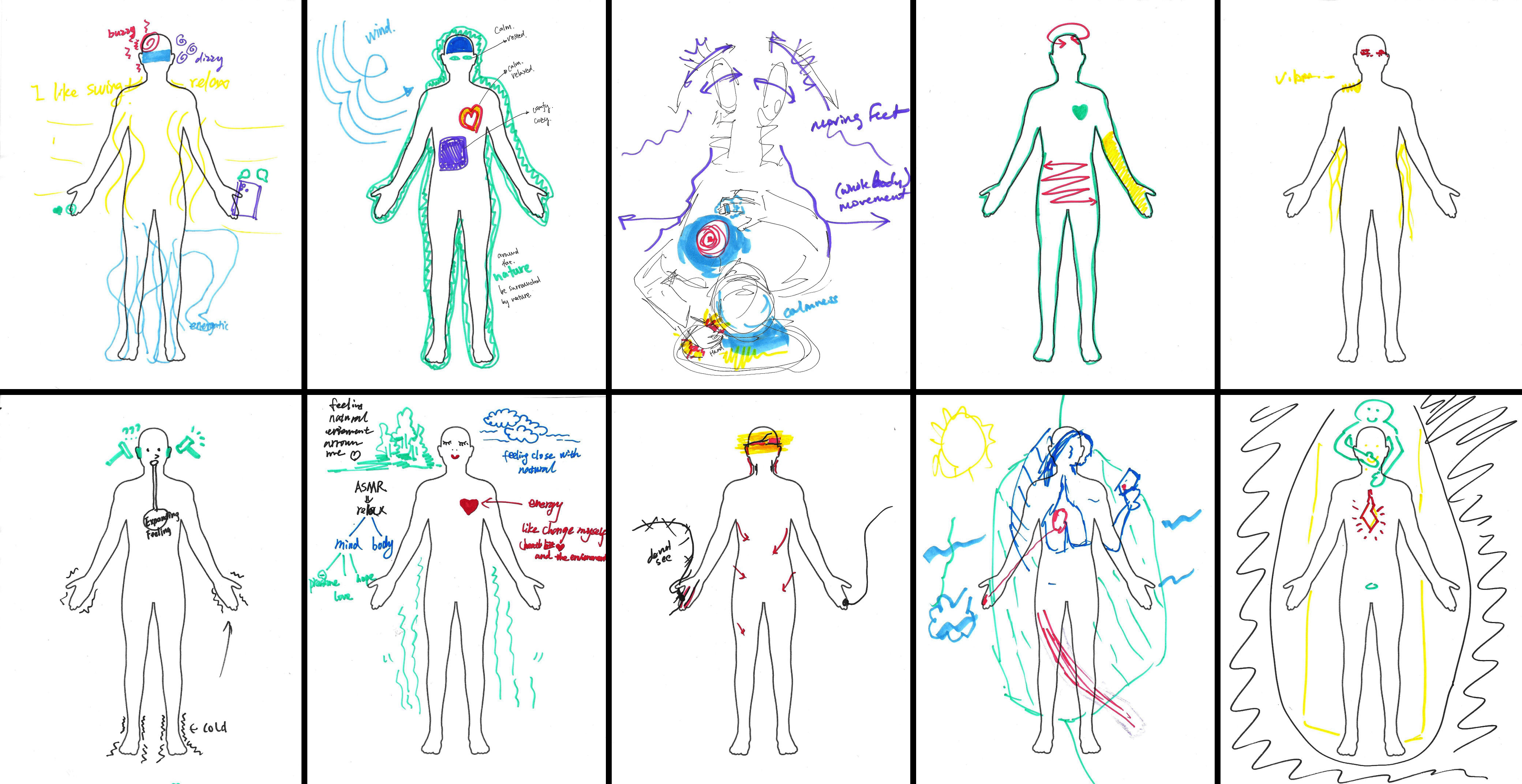}
  \Description{Ten different colorful drawings that depict users' various sensory experiences.}
  \caption{Participants' body maps (A1--A5 in the upper row, A6--A10 in the lower row, from left to right). They are used to elicit more comments on sensory feelings and interpretations during the interview. As depicted above, participants have seen the biodata trace as a spiritual companion, an energetic and inviting subject, a harmonious mirror of the environment, and so on.}\label{fig:body-map}
\end{figure*}

In the semi-structured post-interview, we first inquired about participants' direct personal experiences with this new form of traces in public spaces.
Then, we asked how these personal experiences might shape their impressions and imaginations of the previous and the next person---to probe how the design might evoke reflection on social cues of urban traces that link strangers together~\cite{murphyDwellingTogetherObservable2017}.
The first two aspects were further elaborated as we asked participants to draw body maps~\cite{annecochraneBodyMapsGenerative2022} (\cref{fig:body-map}) and describe them.
Finally, we asked whether participants had any new conception of this campus community after interacting with \name.

To analyze the data, two authors first independently transcribed the audio recordings and cross-checked the correctness.
Then, the same two authors conducted a thematic analysis~\cite{clarkeThematicAnalysis2014} on the interview data.
The authors first independently extracted codes regarding participants' conceptualizations of their experiences and their reflections. Then, the authors jointly discussed the codes to find a dimension to organize them. We decided to particularly focus on different social scales---the self, the triad of consecutive users, and the overall community. With these themes, we revisited the transcripts, revised codes when necessary, and jointly discussed to reach alignment and categorize the codes.

%

\section{Findings}

\edit{
Our findings largely affirm \name's design intent of fostering poetic and authentic asynchronous, person-to-person connections, with participants reporting curiosity about and empathy toward their unknown predecessor. Beyond this, however, the deployment surfaced additional layers of connection that extended beyond our initial goals, including attunement to the surrounding natural environment and associations with intimate personal relationships. Rather than treating these as deviations, we understand them as productive expansions of our design intent that emerged through real-world deployment. We organize our findings across three social scales: participants' direct personal experiences, their conceptualization of the triad of consecutive users, and their broader reflections on community.}

\subsection{Direct Personal Experiences}

This subsection unpacks participants' mix of sensory, emotional, and cognitive experiences. Although some of these experiences were not directly related to the social and empathetic conceptualization of biodata traces, we found them to have crucially established a theme that is peaceful and lively, 
\edit{bodily-oriented}, and socially inviting.

\subsubsection{Relaxation and Comfort}\label{result:personal:relax}

Participants generally acknowledged that the overall sensory and bodily experiences were relaxing, peaceful, and comforting (A2, A4--A5, A7--A10).
The heartbeat vibration is relatively more saliently perceived, with participants commenting that it is reassuring (A2, A4), comforting (A7, A10), and intimate (A10).
The hammock swinging is mainly perceived as a milder experience, relaxing (A3, A4, A5, A9), and reminiscent of parental care in childhood (A1, A9).
Sometimes, the overall experience even led users to a meditative state (A5, A10).
Moreover, participants remarked on the synergy between the haptic features and the hammock's reclined position and the peaceful natural environment (A1, A3, A7, A9).
In the body map, A4 used green borders to depict a state of being ``surrounded'' by safety, and she elaborated, \emph{``The vibration made me feel safe. The movements of this bed also made me feel as if I was in a car or swinging like a baby\dots{} The whole stuff made me feel comfortable, and I almost fell asleep.''}
Meanwhile, A2, who held the vibrating pillow, drew it as a purple patch that generated physical warmth and coziness.
Sometimes, the haptic feedback was also perceived as somatosensory stimuli that actively invites users' attention to their own bodies and minds (A8, A10).
In their body maps, A8 drew black lines to indicate that the experience served as a ``gateway'' of his attention. A10 similarly drew a black ``protective shield'' surrounding his body outline, illustrating how he felt the ``annoying'' urban noise was filtered out in this private, immersive space.

The sensations of vibration and swinging were also found to be playful and interesting (A1, A2, A4, A7, A9), with the vibration sometimes articulated as intriguing (A1, A2, A9) and the swinging fun (A1, A3) and energetic (A1).
With the flexibility of a pillow, A1, A2, and A8 strategically changed their body positions to perceive the heartbeats in different ways.
For A7, this integration of hammocks' relaxation and biodata's vitality \emph{``is a good way to charge my body''}, which is illustrated as green ``energy'' waves around her body.

Thus, the installation fostered a state of tranquil embodiment, where the relaxing nature of the hammock was enlivened by the playful, intimate, and vital presence of the biodata traces.

\subsubsection{Environmental and Physiological Attunement}\label{result:personal:aware}

\edit{Participants described the experience } 
as a positive interruption, drawing attention away from the hustle and bustle and toward a balance between the self and nature (A1--A3, A6--A9).
On the one hand, it was outlined by the theme of biodata and embodiment.
As highlighted by A1, who saw herself as easily captured by the digital realm, \emph{``Every time [the hammock swung]\dots{} it reminded me that I'm here in this environment. Because when I focused on my phone, I would just completely forget the outside world.''}
Also revealed in body maps, A9 drew lungs on her map to mark a newfound attention to her breathing, and A10 used red to signify the sensation of his own ``flesh and blood'' resonating with the stimuli.
On the other hand, the engagement with self and nature was synergistically facilitated by the hammock outdoor setting itself.
As A3 commented, \emph{``[Because of the pillow,] my head was popped up a little bit, and I was looking forward\dots{} out at the trees, which I thought was interesting. I don't always take the time to sit and look at trees. It was very nice to be aware of my surroundings in that way.''}
Participants generally acknowledged a soothing sensation and detachment from daily stresses, which was seen as fundamental for enabling participants to engage more fully with the sensory data transmitted through the hammock (A2, A3, A8).

Overall, the physically grounded context underlined the significance of embodiment during the experience. It has laid an important foundation for participants' reflection on social connections and the community, which is covered in the following subsections.

\subsection{Genuine Connections among the Triad}

This theme explores how participants conceptualized the triad of previous, self, and subsequent users, as well as how a sense of connection, curiosity, and communication was fostered at an individual level. To articulate this theme, we draw from the concept of \emph{mediated genuine connection} by Stepanova et al.~\cite{stepanovaStrategiesFosteringGenuine2022}. Mediated genuine connection describes the felt experience of connections between people, and they offer design strategies for fostering technologically mediated interpersonal connection~\cite{stepanovaStrategiesFosteringGenuine2022}. In the following, we refer to some of these design strategies to help articulate dynamics that emerged with HeartSway regarding how our participants perceived genuine connections through interactive biodata traces.

\subsubsection{Feeling an Intimate Connection}\label{result:triad:intimate}

In describing their experiences, participants used \emph{embodied metaphors} (using physical metaphors for experiences or concepts) and \emph{touch} (sending touch through tangibles)~\cite{stepanovaStrategiesFosteringGenuine2022}.
The former was often denoted as lying and swinging with someone else (A1, A3, A9), and the latter was usually associated with intimate actions like hugging or cuddling (A9, A10).
For example, A10 recalled, \emph{``It makes me think of hugging my mom, I can hear the heartbeat. When we hug anyone, we lie our heads on their chest, and we can feel their heart beating.''} In his body map, he also drew this metaphorical figure embracing his body outline from behind.
In alignment with the vibrancy of heartbeats~\cite{howellLifeAffirmingBiosensingPublic2019}, its preservation and reenactment were not perceived as mere abstract data but as a representation of human life, warmth, and presence, creating intimate connections with the previous user.

Moments of ambiguity---either originating from the form of interactions or their conceptual indications---also occasionally sparked (positively) quirky sensations that raise users' awareness of being ``in the loop'' with the unseen other (A3, A8).
For example, A3 underscored their sense of ``uncertainty'' as \emph{``I have no way of knowing if I'm giving this impression of calm or not\dots{} What message am I even sending?''} Yet, they liked such uncertainty as it distinguished \name from medical scenarios like a hospital bed; and \emph{``It allows me to let go a little bit and not to be so concerned with my own body, but maybe see myself as in communication with these other bodies.''}
In the meantime, A8 remarked, \emph{``I sometimes felt like it was my heartbeats\dots{} where I would confuse why my heartbeats to be like that\dots{} So that was the moments when you were kind of like, \emph{`Why is it racing like that?'}''}

These moments of ambiguity further sparked curiosity and reverie about the previous user (A3, A4, A6, A8, A9) and the future user (A3, A4, A7, A8).
For the former, it was mainly because of the organic fluctuations in the biodata that were usually interpreted to embed some contextual information, as \emph{``if [the vibration] was a uniform rhythm, you start to ignore it after some time''} (A8).
With perceptible variation in the swinging to display the toss-and-turn of the previous user, this prompted participants' open interpretation, such as imagining, \emph{``what kinds of things happened at that time?''} (A2) and \emph{``what the last person was thinking?''} (A6).
Participants were also curious about how the next person might perceive their biodata and their peacefulness (A3, A7, A8).
A7 elaborated, \emph{``I can imagine that when you lie down, now you know the last person's feelings, and you also have your feelings. Even if you don't [verbally] communicate with the last person or know this person, you will feel that this is exactly human life, human warmth.''}
This suggests a sense of one-directional \emph{affective self-disclosure}, which Stepanova et al.\@ also stressed can be induced by biofeedback sharing~\cite{stepanovaStrategiesFosteringGenuine2022}.

This blend of intimate metaphor and affective curiosity thus established a powerful sense of presence within the triad, evoking human warmth and imaginative connection.

\subsubsection{Asynchronous Communication and Consideration for Others}\label{result:triad:comm}

Participants described the interaction as a form of communication, where they could receive a ``message'' from the past and intentionally leave a trace for the future (A3, A6--A10).
Heart rates were more considered symbols of emotions and friendly presence, as A7 put, \emph{``I'm very excited that this installation can pass my feelings to the next person, like a communication.''}
A10 also mentioned having put his hand over his own heart to stay in-the-loop with the rhythm, suggesting that he was engaging in an imagined dialogue during his experience.
The hammock swings, on the other hand, were more considered as abstract messages, and the communication process mattered more.
As A3 commented, \emph{``The other tool that I have to communicate with the next person is movement\dots{} I have much more motor control over my own movements\dots{} so I should probably say something to the next person through this gesture.''}
Because most people have more direct conscious control over their bodily movements than their heart rates, participants more readily considered altering their bodily movements as a `message' for the next person.

The intimacy of sharing biodata also fostered a sense of care and responsibility (A3, A6, A8, A10).
When A3 thought they had observed abnormality in the previous person's heartbeats, they expressed concern for the previous user's well-being: \emph{``I  was left with this experience of feeling kind of powerless to help a person who was not having a very calm experience.''}
This suggests A3 felt concern or perhaps even care for the previous person.
And A8 was also concerned whether he had left a peaceful sensory experience for his next user, \emph{``If someone is getting biodata of mine, and it's very intimate like heartbeats, I want them to be calm\dots{} I hope it brings relaxation to someone.''}
Again, this suggests A8's consideration for another.
And even if participants did not easily emotionally resonate with other people, some also suggested their capability and willingness to show care with their biodata traces.
Exemplified by A6, who said, \emph{``I don't mind or care about sharing my biodata. But if someone comes here to the hammock because they feel lonely, I think that will help.''}

In addition to what we have observed, A8 also envisioned \name to afford extra social potential, like a message board. When many people arrive at it in a row, A8 envisioned how in-person social interactions might naturally emerge: \emph{``something like: \emph{`Hey, I loved your heartbeat. I wanna do this again. Do you wanna sleep 1 hour before me, [so that] I can sleep with your heartbeats again?'}\,''}. It further triggered his reflection on what forms of social relationships people could possibly have:
\emph{``It depends on cultures and societies, but in some countries, strangers don't talk. It is a nice added thought to think about someone you don't know.''}
Interestingly, A8 also ascribed the hammock setting to the reflections above: \emph{``I like the change of perspective\dots{} When you lie down and look up, your view of the world changes, and maybe that influences your view of other people.''}
This suggests \name's hammock setting may help people be receptive to the biodata display.

In sum, \name's unique communicative channel allowed a sense of mutual consideration for others to flourish between otherwise disconnected individuals.

\subsection{Conceptualizing the Community}

This subsection focuses on participants' broader, societal-level reflections on community, touching on feelings such as humanity, loneliness, ephemera, and the nature of public connection.

\subsubsection{Safe Intimacy: Anonymity as a Social Catalyst}\label{result:comm:connection}

\name was also seen as proposing a generally novel and meaningful way to connect strangers in public space.
This mode of interaction from a holistic view could foster appreciation and a sense of shared existence while maintaining a comfortable social distance (A4, A7, A8).
\emph{``It connects more people together, which gives this hammock more meaningful significance,''} said A4.
A7 added that, \emph{``This installation does not force you to make a very close connection or take your personal sensitive information. I still have a good distance with strangers, and I can also connect with people I don't know. I think it's a very smart idea.''}
In particular, the attributes of anonymity and asynchronicity were considered crucial.
As users might have different social preferences and patterns, they appreciated how \name mitigated the potential risks or social pressures of face-to-face interaction (A6, A7, A9, A10), making their embodied communication \emph{safe}.
A6 articulated a more socially reserved perspective, saying, \emph{``I feel better interacting with someone who's not here. If we meet in person, it feels a bit intimidating. So it feels nice to see other people [living] in this world in this way.''}
A10, with candid characteristics, said, \emph{``Sharing myself is a glad thing to me, but I'm a little shy in front of the camera. So this is a very good method\dots{} I think deep in mind, everyone would like to share about themselves. But some are too shy to show up, afraid of being recognized, afraid of being attacked\dots{} I think when you say that they can share their biodata traces [like this], they may be happy to do it.''}
This interpretation layer also echoes what Stepanova et al.\@ coined \emph{provocations}, where connections are established through overcoming social discomfort~\cite{stepanovaStrategiesFosteringGenuine2022}.

Some participants also mentioned that the experience encouraged resonance with the inner lives of strangers, humanizing them and potentially fostering greater acceptance and lower social defensiveness within a community (A8, A9).
For example, A8 stressed that \emph{``It makes you think that there are so many more people that have feelings and heartbeats. And they are also important.''}
Such humanized, authentic depictions of others may be especially valuable when the community is new to the user (A7, A9). As articulated by A9, \emph{``If I come to a new community and I experience something in devices like this, I definitely will become more likely to accept others, feeling that \emph{`other people are also normal human beings once in that hammock'}.''}
As participants gradually gained a sense of affiliation from seeing such commonality, their \emph{reflection on unity} further fostered genuine connection with other strangers in the community, echoing another strategy from Stepanova et al.~\cite{stepanovaStrategiesFosteringGenuine2022}.

\subsubsection{Existential Resonance: Loneliness and Ephemera}\label{sec:existential-resonance}

With the embodied and intimate representation of another person, the biodata traces provided a powerful feeling of not being alone, combating loneliness, and fostering a sense of belonging in the world (A6, A7).
A7 imagined, \emph{``For lonely people, like when living alone or just coming across a new city, this installation offers a positive way to say, \emph{`You are not alone, you actually live in a vibrant world.'} It can make the person feel strong and belong to the community.''}

A3 formed an alternative interpretation of the biodata traces, describing them as ephemeral. They described experiencing only the person immediately prior as a metaphor for the inevitable loss of generational memories.
They found it symbolic of how memories are recorded, fade, and become forgotten over time: \emph{``it feels impossible to really know what somebody's experience is. All we have are these like imperfect tools to represent biodata.''}
While engaging in open interpretation in response to the biodata display, A3 also acknowledges the limits of what could be known through a biodata display.
This conceptual resemblance also triggered their further reverie and fascination, \emph{``it was interesting to realize that you (researchers) have all previous people's biodata, but I couldn't access it. I was only aware of the person right before me, but I kind of had this ghost impression of the collective of all these other people that did it.''}
In that sense, \name accentuated the eventual erosion of memorabilia and memories, inviting people to cherish the ephemeral presence.

\section{Discussion}

Based on our design process and findings, here we further articulate our designerly reflections that emerged throughout the entire project process~\cite{sengersReflectiveDesign2005}. These higher-level insights gathered from the design and evaluation of \name may inform future HCI researchers on urban traces and biodata.

\subsection{The Materiality of Biodata as Archive}

With \name, we explore an under-examined dimension of designing with biodata: \textbf{archiving biodata as traces for asynchronous perception by strangers.}
Our initial motivation drew an analogy between biodata recordings and poetically ambiguous urban traces, such as footprint impressions and warmth left on a public bench. 
With our findings, this section revisits this analogy and reflects on the relation between archival biodata and its sense of materiality.
Specifically, we put forward the following three aspects for possible future research directions: biodata datasets as archives embodying affective messages, as a catalyst for poetic reverie, and as a medium possessing inherent ephemerality and imperfection.

\subsubsection{Biodata Datasets as Tangible Affective Archives}

The asynchronous interaction in \name may have reinforced the sense that biodata is treated as a material archive of another person's affective state.
``Affect is a force that creates a relation between a body and the world'' \cite[p. 8]{ciforAffectingRelationsIntroducing2016}, and \name prompted participants' appreciation of relations between their body, their surroundings, and a past person's presence.
Taking a step further, we underscore three characteristics of our archival design pattern that relate to materiality.

Users are invited to effortfully interpret affective meaning from the data's encoded form.
This pattern foregrounds the biodata's role as \textbf{material carriers} and \textbf{interpretative substrates} that embody communicative affordances~\cite{liExploringOpportunitiesAR2023,horneckerDesignVocabularyData2023}.
This extends a legacy of HCI design research (e.g., \cite{boehnerAffectInformationInteraction2005}) that, as alternatives to typical approaches in affective computing or Emotion AI aiming to computationally interpret emotion from biodata, instead centers human interpretation of emotion and the relationality of affect.
When designs direct users' attention toward the biodata itself and refrain from interpreting it for users, they may help exhibit biodata's inherent expressivity about human affects and vibrancy.

Second, this materiality could also be associated with \name's tangible design, which approximated the original somatic forms~\cite{alfarasBiodataSomadata2020} of the biodata (heartbeats, body movements).
In doing so, it shaped what has been termed ``tangible affect'', reflecting materiality from the \textbf{process of physicalizing the intangible affects}~\cite{zhouTangibleAffectLiterature2024}.
\edit{Also echoed in participants' comments, the physical forms could intuitively preserve and vividly tell human physiology behind the biodata~\cite{zhouTangibleAffectLiterature2024}.}

Lastly, \name's use of biodata as \textbf{archival} also sharpened its materiality.
This aligns with Riggs et al.'s research~\cite{riggsDesigningArchiveFeelings2024}, in which they challenge the traditional notion of archives as scientifically neutral historical records.
They show that tangible and affective archive materials effectively prompt reflection and personal meaning-making around community history~\cite{riggsDesigningArchiveFeelings2024}.
\name also uses biodata in tangible and affective archives, shaping similar materiality that makes history feel personal, immediate, and emotionally alive.

\subsubsection{Biodata as a Catalyst for Poetic Reverie}

Beyond embedding information and engaging users in deciphering actions, the biodata traces in \name have \textbf{guided participants into layered reveries} about human affects both from the past and in the future.
In this sense, these biodata traces were not only recordings but also evocative materials for reminiscence.
The biodata's existence indicates temporal links between previous, current, and future users, raising participants' awareness of their place in connection to others, part of a continuing chain of hammock visitors in a shared public space.

Previous literature also suggests that to advocate embodied, and performative understanding of affect, designers should favor ``infusing affects into designs'' over ``embedding affective information''~\cite{kozelAffeXityPerformingAffect2012}, and \name is consistent with this idea.
Rather than serving information, \name infuses a sense of connective forces between each participant's body and the linked chain of hammock visitors.
To better articulate this, we draw from Stewart's \emph{ordinary affects}, which describes objects' affective capability to affect and be affected that gives our life the quality of contingence~\cite{stewartOrdinaryAffects2007}. Stewart further stresses the poetic potency of \emph{ordinary affects} among \emph{``found objects snatched off the literal or metaphorical side of the road\dots{} as if they are literal residues of past dreaming practices''}~\cite{stewartOrdinaryAffects2007}.
Thus, \name suggests the potential for biodata traces to become poetic memorial material evoking \emph{ordinary affects}.


\subsubsection{Biodata Ephemerality and Imperfection}
Finally, we consider the \textbf{inherent ephemerality and imprecision} of biodata archives as another aspect of their materiality.
\edit{Ephemerality here operates on two levels: biodata signals are fleeting by nature, and by design, each session's data is discarded after use, ensuring users encounter only the trace of their immediate predecessor.}
The chained process, where a previous user's data is overwritten and thus lost, reminded some participants of how physical memorabilia wear out and histories fade.
This resonates with Gulotta et al.'s critical design where digital assets are made to decay to heighten their owners' awareness of their value~\cite{gulottaDigitalArtifactsLegacy2013}, or Seznec's destruction of personally meaningful audio recordings~\cite{seznecThreeWaysDestroy2024}.
Our user study suggests that occasional noisy sensor data could sometimes reflect the inevitable inexactitude in materializing intangible histories and human expression.
In this way, biodata archives may become more evocative as they manifest the ephemeral and flawed nature of physical materials, gently nudging users to cherish the transient perceptions afforded by the traces they encounter.

\subsection{Users as Soma Experience Creators}

When users interact with \name, not only are they creating traces from a material perspective discussed above, but they are also assigned the role of an ``unwitting but potent'' \textbf{creator of soma experiences}.
\name explores potential for participants to craft soma experiences for others, transforming the ostensibly passive state of resting into a generative act: authoring a future, felt experience for an anonymous other.
Here, the creative agent is not the cognitive mind issuing commands, but the largely involuntary soma itself.
As echoed in our user study, some participants were consciously aware that their physiology was shaping a future experience and developed a mindful attention towards this unique, unverifiable act of creation.
This ``somatic authoring'' mode suggests considerations for future investigations.

Soma design advocates for designing with and for the lived, felt body, appreciating its aesthetic and experiential capacities~\cite{hookDesigningBodySomaesthetic2018}.
With \name, as users are engaged in a somaesthetic creation activity for others with their own soma, \textbf{the process itself prompts a deeper introspection of their own lived, felt body}.
Furthermore, as the agency is relocated from intentional action to passive presence, the system prioritizes the user's state of \emph{being} over their active manipulation, consequently revealing the implicit aesthetic and experiential qualities of human bodily presence.
While \name diverges from typical soma designs (e.g.~\cite{dublinJourneyInwardSomaesthetic2024,daudenroquetInteroceptiveInteractionEmbodied2021,jonssonAestheticsHeatGuiding2016,karpashevichTouchingOurBreathing2022}) where the system directly amplifies or guides users' inner states, \name offers a related design exploration of a linked chain of somatic experiences across strangers.

Secondly, users' authoring process is never fully autonomous, but profoundly interdependent.
With the current user continuously perceiving the previous user's physiology, the process can be seen as an \textbf{asynchronous, bodily duet}: the first user provides a ``somatic score'' through their archived biodata. The second user's body becomes the ``instrument'' that performs this score, with their own physical and emotional state mediating the perception of the stimuli.
This establishes a form of technologically mediated intersubjectivity, where the sensory experience becomes not only the signal of the immediate predecessor but also a shared bodily medium~\cite{merleau-pontyPhenomenologyPerceptionIntroduction2002} connecting a lineage of previous inhabitants.

Finally, \name \textbf{reframes authorship by shifting creative agency from direct conscious action to less directly intentional bodily processes} such as heart rates. By transforming passive presence into an embodied dialogue between strangers, it may open new possibilities for designing asynchronous, somaesthetic, and social experiences.

%
%

\subsection{Biodata Traces as Part of the Place}

In this section, we delve deeper into the attribute that \name anchors biodata series in a public place. We present design considerations for future designers' reference, as well as possible future research directions.

First, \name demonstrates \textbf{reappropriation of traditionally single-user public amenities for multi-user, asynchronous interaction}---particularly those that encourage comfort and lingering.
All these amenities are implicitly associated with the infinite array of users that have interacted with them at different times, and place users in a relaxed, open, and unguarded state.
This way, their shared spatial, behavioral, and mental context creates a fertile ground for genuine connections, positive placemaking, and spatially-grounded synchrony.


Second, this work may present another research direction of \textbf{placemaking potential} of biodata traces.
Although the limited scale of our current study cannot support such a generalized claim, we do observe participants' articulations of how biodata traces were perceived as a testimonial to the community's positivity and welcoming atmosphere (\cref{result:comm:connection}).
This remark aligns with Stal et al.'s finding that manifestations of personal affect can enrich urban environments and foster place attachment~\cite{stalsExploringPeoplesEmotional2017}.
And since biodata traces ground individuals in their own corporeality while revealing the ``human warmth'' of co-occupants, they also appear to align with Gustafson's experiential and social dimensions of place~\cite{gustafsonMEANINGSPLACEEVERYDAY2001}.
With more design explorations and extensive studies, future work can continue to explore how biodata traces may facilitate urban placemaking.

\edit{
Moreover, \name does not simply broadcast information to an already overburdened urban society. Rather, it deliberately restrains its design with one person at a time, tactile rather than visual, passive reception, and a restful bodily state.
We realize that this slow and focused mode of connection---originated from the analogy to traces---also resonates with the vision of \textbf{calm technology} in HCI: information can be moved to the periphery of attention until needed, rather than demanding constant engagement\cite{WeiserDesigningCalmTechnology1996}.
Although this paper did not particularly evaluate \name from the calm technology perspective, we believe it is an interesting future research direction to go deeper into the link between public trace designs and calm technology.
}

Finally, an interesting part of our findings is that some users find it evocative to enjoy the combination of someone's authentic life signals and the relaxing natural environment of trees, ambient sounds, and the sky (\cref{result:personal:aware}).
This \edit{reminds us of the notion of} \textbf{more-than-human}, which \edit{spotlights the agency and vibrancy of non-human objects in our everyday surroundings}~\cite{janickiCripReflectionsDesigning2024,oferTracingStrategyOrienting2024,sondergaardFeministPosthumanistDesign2023}.
\edit{Although biodata traces in \name are produced by human objects, it coincidentally helps us realize this potential underlying link between traces and more-than-human designs.
We believe that future designs surrounding non-human objects can further explore this link to effectively deliver positive messages to the place.
For example, designs surrounding plants' and animals' biodata at a place of interest may both acknowledge the ephemeral and negligible beauty and the agency of those non-human lives, contributing to residents' knowledge and impressions of the place.
}

\subsection{Tension between Public and Private}

By embedding internal biosignals into shared environments, \name introduces tension regarding privacy boundaries in urban spaces. Our user study suggests that this tension is not inherently negative, but it renders the experience provocative and generative. Here, we further reflect on this tension and consider design recommendations around boundary-respecting systems.

\edit{The first point is \textbf{involuntary public disclosure}.}
\edit{This} is enforced when \name shifts control of disclosure away from users, despite biosignals being inherently personal.
Since sharing biodata is not a common practice in participants' daily lives, this unusual experience has prompted them to reflect on the societal value of their biodata, as well as optimal formats and scenarios for meaningful sharing.
Similarly, Howell et al.'s previous work also suggested tension surrounding biodata interpretation when it is ambiguously and non-authoritatively displayed to the public~\cite{howellTensionsDataDrivenReflection2018}.

In our case, where empathetic connections are particularly desired, we suggest data anonymization and abstraction. \name deliberately decouples the data from user identity and translates precise metrics into non-numerical haptic signals.
This may help prevent the exposure from becoming surveillant or invasive, yet still encourage reflection on sharing practices.
\edit{But still, participants have highlighted a deeper bias that we had neglected before.
As A3 mentioned, the designers have all the data while users only sense the previous person's and ``send'' their own (\cref{sec:existential-resonance}).
This power imbalance may be seen by some future users as a hindrance to a truly comfortable disclosure.
Therefore, designers can seek structurally alternative ways to handle data and keep users fully informed of the data policy.}

Secondly, \name also advocates \textbf{social connections} by compelling users to physically feel the vitality of an unknown stranger.
Hammocks may typically serve as private retreats for solitude and detachment from the world.
But with the biodata naturally embedding affects, its communication is consequently prosocial~\cite{nunezEffectSocialConnectedness2019}. It nudges users to overcome \emph{civil inattention}~\cite{hirschauerDoingBeingStranger2005} and prompts new forms of engagement with strangers.
To ethically imbue social implications, designers should particularly minimize social risks and awkwardness.
For instance, \name only communicates non-verbal biodata to prevent toxic exchanges and over-sharing, and instead offers a simple, low-effort way to broach anonymous connections.
\edit{Meanwhile, we must also note that our design of reappropriating hammocks is tied to the local context, such as the vibrant campus atmosphere and prior familiarity.
With other specific cultural values and contexts, designers can not only rethink the strategies of prompting strangers' engagement, but also the level of ``good distance'' to preserve with the design.}


\subsection{Limitations and Future Work}

Our user study was deployed on a university campus and \edit{only covered} university students. The \edit{lack of other identities among campus residents, such as service staff and professors, the} small sample, \edit{and the limited coverage of user backgrounds} may not fully reflect the general public's attitudes towards the design.
The one-time nature of our study may additionally limit our understanding of how users might engage with \name over extended periods. Thus, future long-term deployments may provide insights into whether users develop different interpretations or interactions with the system over time, and whether serendipitous~\cite{almqvistDifferentTogetherDesign2023} or ritualistic~\cite{leongSocialWormholesExploring2023} uses emerge.

As mentioned in \cref{sec:design:implementation}, \name's implementation was simplified during our real-world deployment and user study, because we needed to repetitively install and remove the system if we wanted to deploy and test the design on existing hammocks in a public area. \edit{While the Wizard-of-Oz approach is practical, it carries additional limitations. First, the exact timing of body-moving movements can not be precisely replicated manually, potentially reducing the fidelity of the embodied experience. Second, the physical presence of the researcher operating the string may have influenced participants' perception of the swinging, potentially enlarging the playful attribute of the experience.}
In the future, we will seek opportunities for continuous and unstaffed deployments, \edit{potentially with portable 12V car batteries}. These tests can help us investigate potential real-world challenges, such as weather resilience and maintenance.
\edit{Additionally, future versions can implement more detailed physiological tracking, such as the intensity of tossing and turning, to offer a more realistic embodied experience.}


\edit{Finally, we hope the design knowledge generated by \name{} would inspire future explorations across different materials, biodata types, communities, and cultural contexts.
Following the generic idea of preserving and revisiting data series, other types of biodata can be designed to evoke a similar sense of anonymous intimacy and shared vitality, while spotlighting different aspects of human physiology. 
When adapting \name{}'s idea to other communities and cultural contexts, designers may also seek particular materials and information with special affordances to the target audience. It is also worth studying whether and how these special design materials may provide a synergetic effect on the user experience.
More broadly, practitioners in community art, urban design, and public health may find value in celebrating implicit neighborhood togetherness and affording breathable stranger intimacy.}
\section{Conclusion}

In this work, we have explored a compelling design space where physiological data serves as poetic urban traces, creating asynchronous connections between strangers in public spaces.
We designed and implemented a design probe, \name, an interactive hammock installation that captures users' heart rate and body movements, transforming them into haptic experiences for subsequent users.
This probe revealed diverse interpretations of physiological traces \edit{that match or extend our design goals}---from spontaneous empathy and curiosity about previous users to internalization of others' physiological states and poetic appreciation of human presence.

\name contributes to \edit{the design knowledge of} how urban traces can foster subtle yet meaningful human connections in public spaces: the acceptance of, resonance with, and care for unknown strangers in the same community.
In addition, it sheds light on the affective use of biodata in HCI designs.
It demonstrates that biodata, when transformed into ambient, embodied experiences, can enable meaningful socio-emotional interactions in urban space.
It also demonstrates that decontextualized biodata archives can support empathetic social connections without requiring synchronous interaction.

\edit{Beyond the concrete design of \name, the general insights also} open new possibilities for enriching public spaces with poetic, person-to-person interactions that celebrate human diversity and presence without disrupting the natural flow of urban life.


\begin{acks}
     This material is based upon work supported by US National Science Foundation (NSF) award 2335974.
\end{acks}

\bibliographystyle{ACM-Reference-Format}
\bibliography{main}

\appendix

\end{document}